\documentclass[graybox]{svmult}
\usepackage{mathptmx}       
\usepackage{helvet}         
\usepackage{courier}        
\usepackage{type1cm}        
\usepackage{makeidx}         
\usepackage{graphicx}        
\usepackage{multicol}        
\usepackage[bottom]{footmisc}
\newcommand{\ha}{\relax \ifmmode {\mbox H}\alpha\else H$\alpha$\fi}
\newcommand{\nii}{\relax \ifmmode {\mbox N\,{\scshape ii}}\else N\,{\scshape ii}\fi}
\newcommand{\oii}{\relax \ifmmode {\mbox O\,{\scshape ii}}\else O\,{\scshape ii}\fi}
\newcommand{\oiii}{\relax \ifmmode {\mbox O\,{\scshape iii}}\else O\,{\scshape 
iii}\fi}
\newcommand{\sii}{\relax \ifmmode {\mbox S\,{\scshape ii}}\else S\,{\scshape ii}\fi}
\newcommand{\hb}{\relax \ifmmode {\mbox H}\beta\else H$\beta$\fi}
\usepackage{color}
\newfont{\fcm}{cmssdc10 scaled 900}
\let\oldmarginpar\marginpar
\renewcommand\marginpar[1]{\-\oldmarginpar[\raggedleft\footnotesize\color{red} #1\color{black}]%
{\raggedright\footnotesize #1}}

\makeindex             
\begin{document}

\title*{Unveiling the nature of the ``Green Pea'' galaxies}
\author{R.~O. Amor\'in, J.~M. V\'ilchez and E. P\'erez-Montero}
\institute{R.~O. Amor\'in (\email{amorin@iaa.es}); J.~M. V\'ilchez (\email{jvm@iaa.es}); E. P\'erez-Montero (\email{epm@iaa.es}) \at Instituto de Astrof\'isica de Andaluc\'ia, Glorieta de la Astronom\'ia S/N, Granada, Spain 
}
%
%
\maketitle

\abstract{We review recent results on the oxygen and nitrogen
  chemical abundances in extremely compact, low-mass starburst
  galaxies at redshifts between 0.1-0.3 recently named to as ``Green
  Pea'' galaxies.  These galaxies are genuine metal-poor galaxies
  ($\sim$ one fifth solar in median) with N/O ratios unusually high
  for galaxies of the same metallicity. In combination with their
  known general properties, i.e., size, stellar mass and
  star-formation rate, these findings suggest that these objects
  could be experiencing a short and extreme phase in their evolution. 
  The possible action of both recent and 
  massive inflow of gas, as well as stellar feedback mechanisms are 
  discussed here as main drivers of the starburst activity and their 
  oxygen and nitrogen abundances.}

\section{Introduction}
\label{sec:1}
There is strong evidence that compact low-mass starburst galaxies with
properties similar to those of Blue Compact Galaxies (BCGs) have
provided $\sim$40 \% of the star formation rate (SFR) density at
redshift $z = 1$ [1]. The idea of a strong evolution of low-mass
galaxies at a relatively late cosmic epoch (i.e., by the time when the
Universe has been half of the present age) is consistent with
``downsizing'' scenarios of galaxy formation, predicting a strong
evolution with specific SFR progressing from high- to low-mass systems
with time [2].  The rapid decline of these proto-BCGs from $z = 1$ to
$z = 0$ and subsequent evolution are barely understood.

One of the largest and most homogeneous samples of low-mass starbursts
at redshift $z < 1$, is the commonly referred to as ``Green Pea'' (GP)
galaxies, recently discovered by their unusual color and compact size
(unresolved on SDSS images) by the ``Galaxy Zoo'' project [3], and
further studied in some detail by [4], [5], and [6].  The GPs are
mainly located in lower-density environments and are spectroscopically
characterized by very faint continuum emission and strong optical
emission lines, such as the \mbox{[O{\sc iii}] $\lambda$5007} emission
line, with an unusually large equivalent width of up to
$\sim$2000\ \AA.  According to the above studies, at least 80 of these
newly discovered objects are very compact (half-light radius
$\sim$1-2 kpc) low-mass
(M$_{\star}$$\sim$10$^{8}-$10$^{10}$M$_{\odot}$) starbursting systems
(SFRs up to 60 M$_{\odot}$ yr$^{-1}$, as derived from their H$\alpha$
and UV luminosities).  The above properties led to the conclusion that
the GPs are identifiable with the high-luminosity end of nearby BCGs,
probably representing earlier and extreme stages of BCG evolution.
Moreover, their extremely high specific SFRs 
(10$^{-9}-$10$^{-7}$ yr$^{-1}$) are among the highest known in the
nearby Universe, and comparable to those of high redshift galaxies.
The similarities in these properties, and also -- as we will see below
-- those found in their chemical abundances and extinction suggest
physical conditions that largely resemble those in galaxies in the
early Universe.  Therefore, the GPs may be used as nearby proxies for
studying in much greater detail the stellar mass growth of galaxies
and the physical processes involved in the triggering of violent star
formation and the chemical evolution of higher-redshift galaxies.
 
\section{Oxygen and nitrogen chemical abundances and scaling relations}
\label{sec:2}

In [5] we used SDSS spectra to carry out a detailed study of the
chemical abundances of the sample of GPs in the redshift range $0.11
\leq z \leq 0.36$ that were spectroscopically classified as
star-forming systems by [4].  We measured emission-line integrated
fluxes for all galaxies after a linear subtraction of the continuum
and then we computed their oxygen total abundances and
nitrogen-to-oxygen ratios, N/O, using empirical methods (N2 $\equiv$
\mbox{log([\nii] $\lambda$6584/\ha)} and N2S2$\equiv$[\nii]/[\sii]
     [7]).  Additionally, for a subset of galaxies with reliable
     measurements of the [\oiii] $\lambda$4363 emission line (70\% of
     the sample) physical properties and oxygen and nitrogen ionic
     abundances were calculated using the more reliable direct method.

In contrast to [4], our results showed that the GPs are a population
of genuinely metal-poor galaxies, with oxygen abundances $7.7 \leq$
12$+$log(O/H)$\leq 8.4$ (mean value of 8.05$\pm$0.14, or $\sim$20\%
the solar value). This value is $\sim$0.65 dex lower than that
previously obtained by [4] using an empirical calibration based on the
[\nii]/[\oii] ratio.  This large difference was attributed by [5] to
the dependence of [\nii]/[\oii] on the variation of the N/O ratio at a
given O/H, since large values of N/O can enhance [\nii]/[\oii] even in
the low-metallicity ragime [7].  As we shall show below, the N/O
values displayed by the GPs are higher than expected for their
metallicity.  It is worth noting that our direct and empirical
estimates of the oxygen and nitrogen abundances agree well within the
typical uncertainties ($\sim$0.1 dex).  The GPs were also found to
show relatively low extinction, with $C($\hb$)$ and \ha/\hb\ mean
values 0.23 and 3.3, respectively. This confirms that these objects
should be relatively dust free.

\begin{figure}[t]
\sidecaption[t]
\includegraphics[scale=.45]{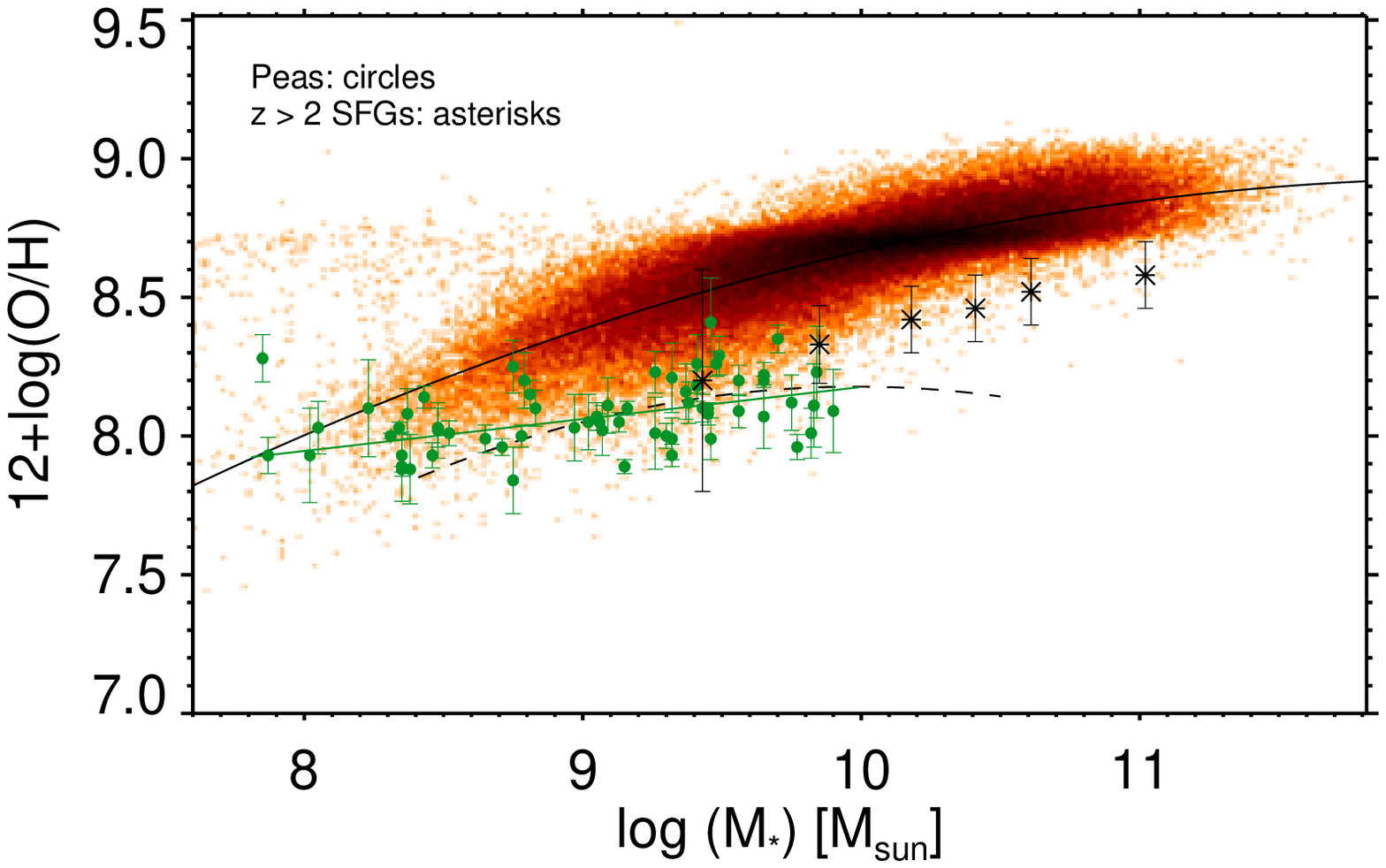}
%
%
\caption{O/H vs. stellar mass. The 2D histogram of SDSS SFGs is shown in
  logarithmic scale and their best likelihood fit is shown by a black solid 
line. 
The subset of 62 GPs (see text) are indicated by circles and their best 
linear fit is shown by a dashed line. 
For comparison we also show the quadratic fit presented in [5] for the 
full sample of 80 GPs. 
SFGs at $z \geq 2$ by [9] are also shown by asterisks for comparison.
}
\label{fig:2}       
\end{figure}

The above results become even more interesting when compared with
those of a statistically significant sample of star-forming galaxies
(SFGs, taken from the MPA/JHU Data catalog of the SDSS DR~7), for
which we have applied the same methods and calibrations. 
For both GPs and SDSS SFGs, we plot the relation between N/O and
metallicity, the mass-metallicity relation (MZR) and the relation
between N/O and the stellar mass in Fig.~\ref{fig:1}, Fig.~\ref{fig:2}
and Fig.~\ref{fig:3}, respectively. The two main results apparent from
these figures will be the subject of our next discussion:


%
\begin{figure}[t]
\sidecaption[t]
\includegraphics[scale=.4]{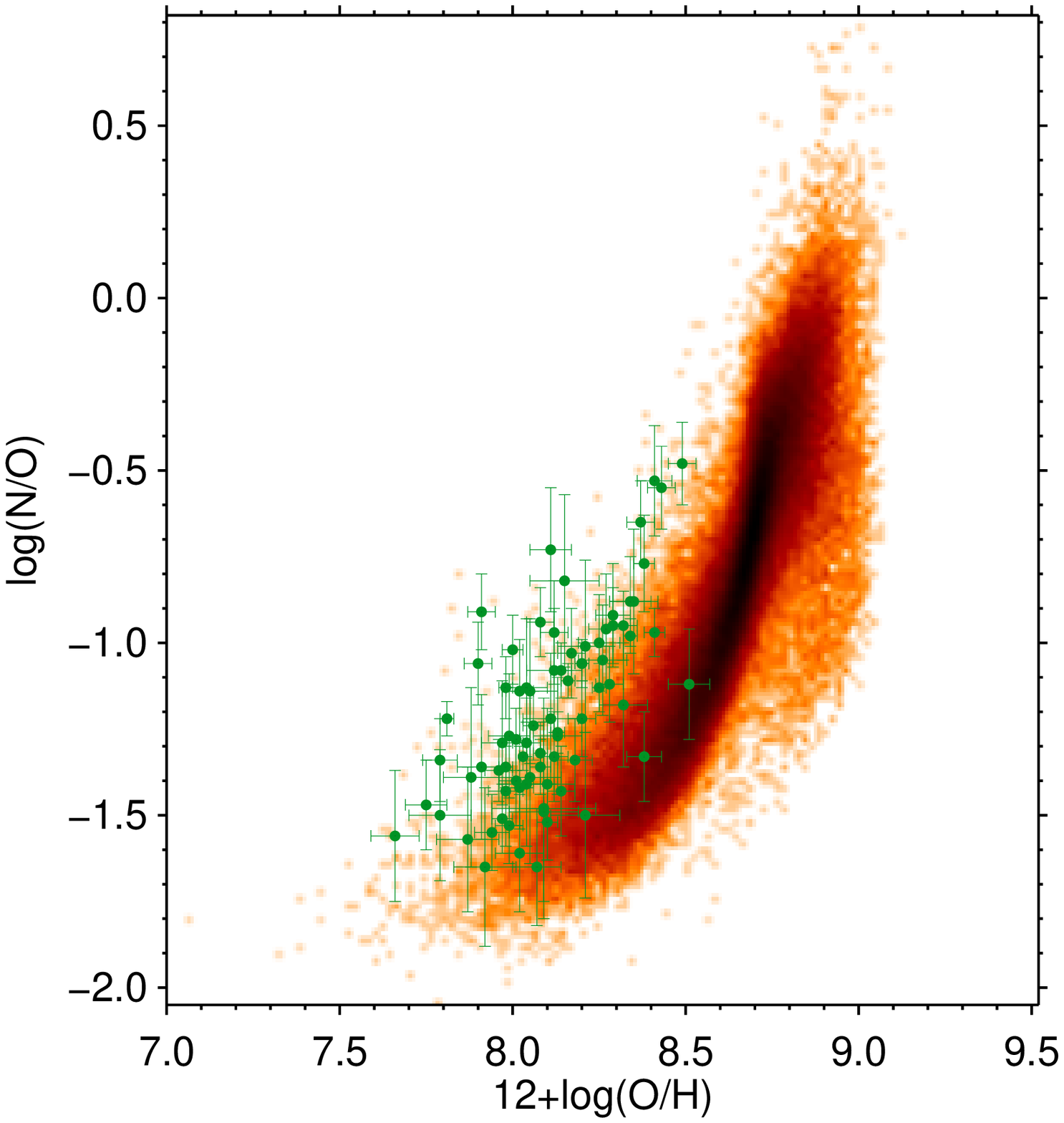}
%
%
\caption{N/O vs. O/H abundance ratio. The 2D histogram of SDSS SFGs is shown in
  logarithmic scale while the GPs are indicated by circles.}
\label{fig:1}       
\end{figure}
\begin{figure}[ht]
\sidecaption[t]
\includegraphics[scale=.45]{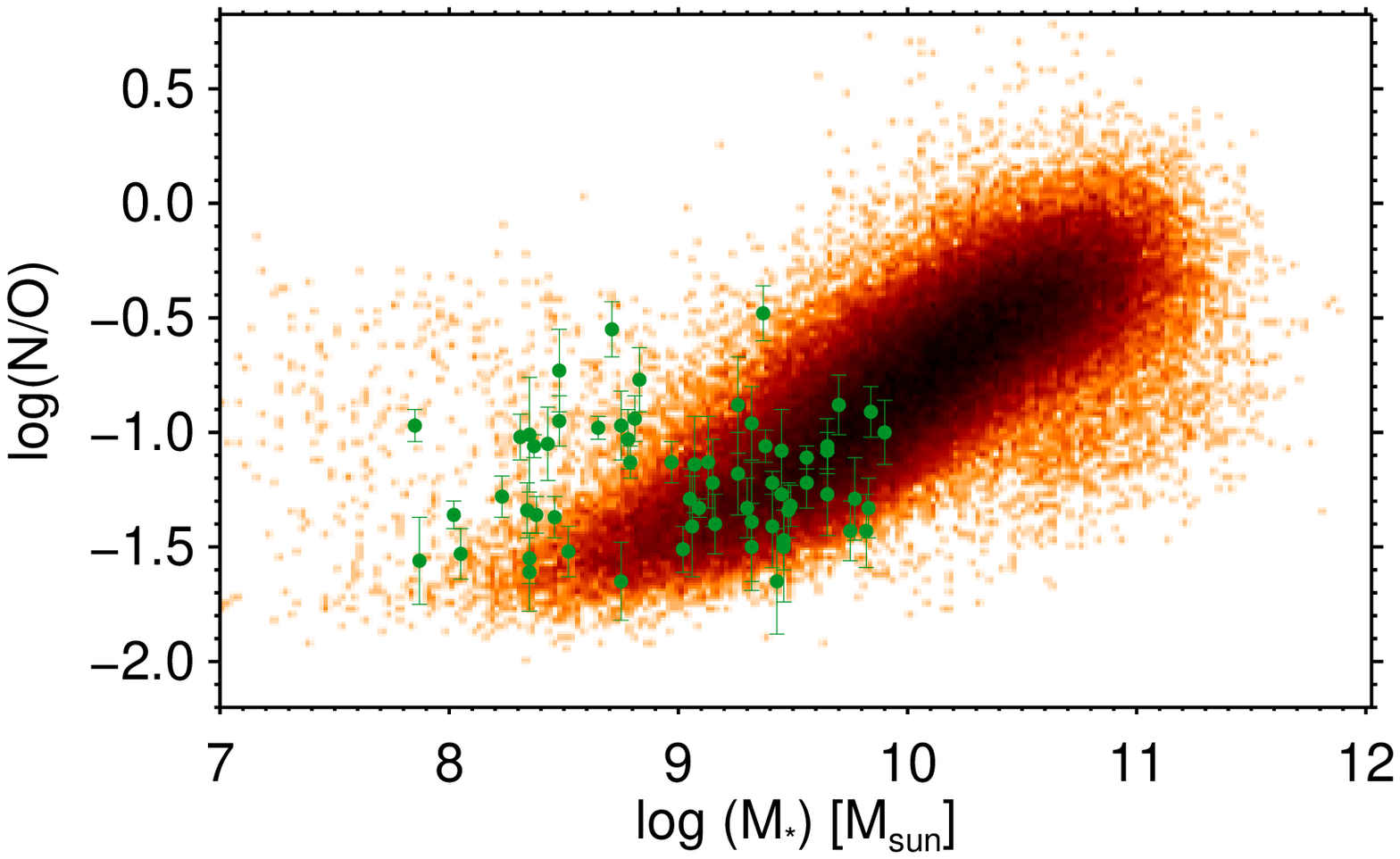}
%
%
\caption{N/O vs. stellar mass. Symbols are the same as in
  Fig.~\ref{fig:1}.}
\label{fig:3}       
\end{figure}
\begin{figure}[ht]
\sidecaption[t]
\includegraphics[scale=.4]{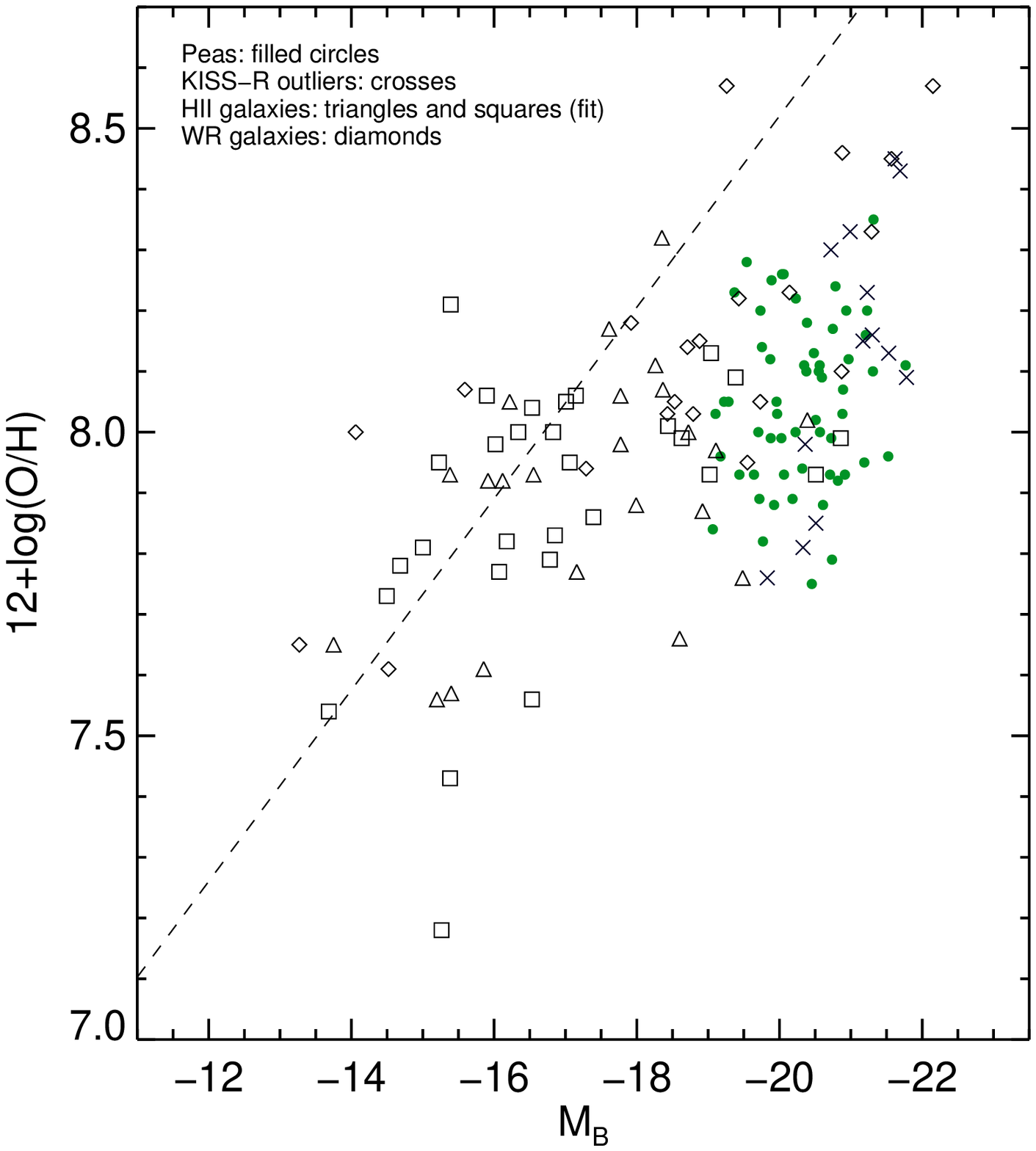}
%
%
\caption{O/H vs. B-band (rest-frame) absolute magnitude. 
The meaning of symbols is indicated. Distances used in computing
(extinction-corrected) absolute magnitudes were, in all cases, 
calculated using spectroscopic redshifts and the same cosmological 
parameters. The dashed line indicates the fit to the H{\sc ii} 
galaxies in the MLR given by [12].}
\label{fig:4}       
\end{figure}

(1) in the low-to-intermediate metallicity range most GPs present
systematically larger N/O ratios for a given metallicity compared to
the SDSS SFGs (Fig.~\ref{fig:1}).

(2) while most GPs show N/O ratios roughly consistent with the
relation with stellar mass of local SFGs (Fig.~\ref{fig:3}), we find
that the MZR of the GPs is systematically shifted to lower
metallicities (Fig.~\ref{fig:2}).

Figures \ref{fig:2} and \ref{fig:3} are similar to those presented in
[5]. However, instead of using $M_{\star}$ derived by [4] (i.e.,
fitting the spectral energy distribution from SDSS photometry after
subtracting the contribution of emission lines), here we have used a
more reliable estimation of $M_{\star}$ recently obtained by [6]. For
62 out of 80 GPs, they derived $M_{\star}$ from fitting their SDSS
spectra after subtracting the contribution of both the nebular
continuum emission and the line emission.  On average, these new 
stellar mass estimates are by $\sim$0.35 dex lower than that of
[4].  However, a comparison of our previous fit in [5] with our fit
using the new $M_{\star}$ values in Fig.~\ref{fig:2} (dashed and solid
line, respectively) shows that the original overstimation of
$M_{\star}$ is not large enough to explain the offset between GPs and
SDSS SFGs.  Only in the lower mass (luminosity) regime, where galaxies
have SDSS spectra with much poorer S/N and therefore are
subject of larger uncertanties in the spectral fitting, the GPs seem
to match the trend of the local SDSS SFGs.

According to their main properties 
(i.e., stellar mass, size, SFR, metallicity) the
GPs appear to bear close resemblance to other, smaller samples of
strongly  star forming galaxies at similar redshifts selected using 
different criteria. 
 That is the case for some Lyman Break Analogs (LBAs) included in the 
analysis done by [5]. They are a subset of 30 local UV-luminous galaxies 
studied in detail by [8] (and references therein) which show strong 
similarities with the more distant Lyman Break Galaxies (LBGs), 
like those at $z>2.2$ studied by [9] that are shown for comparison in 
Fig.~\ref{fig:2}.
The low-metallicity ``Ultra Strong Emission Line'' galaxies found at 
redshift $z < 1$ by [10] or the luminous starburst galaxies at redshifts 
in the range 0.29-0.42 from the KISS survey discovered by [11], also 
show strong similarities with the GPs. 
Interestingly, the latter have been presented as extreme 
outliers in the mass-luminosity relation (MLR) as compared with the 
local Blue Compact/ H{\sc ii} galaxies in the KISS survey.  
In Fig.~\ref{fig:4} we show their position and that of the GPs in the
MLR.  For comparison, we also added a sample of H{\sc ii} galaxies
from [12] and a sample of highly interacting,
local Wolf-Rayet (WR) galaxies by [13].  
In Fig.~\ref{fig:4}, the GPs and the KISS outliers do not
follow the linear relation established by the H{\sc ii} galaxies at
lower luminosities, showing a clear offset of $\sim$2-3 magnitudes. 
This difference in the luminosity-metallicity relation between strongly 
star-forming galaxies (including the sample of GPs) and local emission-line 
galaxies has been also shown by [6], who pointed out that the observed 
trend for GP-like objects is more similar to the one described by  
distant LBGs.  
 
\subsection{Comparison with the closed-box model}

\begin{figure}[t]
\sidecaption[t]
\includegraphics[scale=.4]{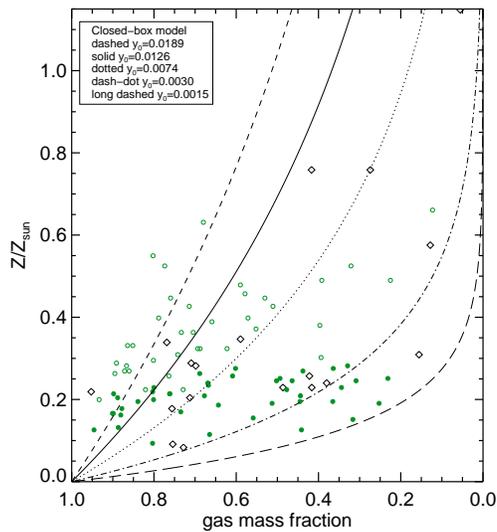}
%
%
\caption{Gas mass fraction vs. metallicity.  Different lines correspond
  to closed-box models at different yields, as indicated in the
  legend. Open and filled circles are GPs which are above and
  below the fit to their MZR in [5], respectively. Diamonds are values
  for the same WR galaxies as in Fig.~\ref{fig:4}.}
\label{fig:5}       
\end{figure}
 
As a preliminary approach in studying our results in terms of chemical
evolutionary models, in Fig.~\ref{fig:5} we compare the observed
oxygen abundance of the GPs as a function of the gas mass fraction
with the predictions of the simplest, closed-box model, that assumes
the galaxies as closed systems, with no inflows or outflows, in addition to
instantaneous recycling and a constant SFR.  In this model the
metallicity is written as $Z = y_{0}$ ln$(1/\mu)$, were $y_{0}$ is the
true yield by mass, and $\mu$ is the gas mass fraction. Since no H{\sc
  i} observations are available for the GPs, the latter has to be
calculated by assuming a star formation law. For doing so, we combined
the correlation between surface gas density and SFR per unit area with
the conversion from H$\alpha$ luminosity to SFR from the relations
given by [14]. Since the GPs are unresolved in SDSS, 
in this approach we considered the area where neutral
gas (involved in star formation) is located as the one defined by the
size of the SDSS fiber (i.e., 3$''$).  In Fig.~\ref{fig:5} the lines
indicate predictions of the model using different yields, were the
black line represent a model with $y = 1 Z_{\odot}$, while the red one
is the theoretical oxygen yield expected for stars with rotation of
[15], $y \sim 0.6 Z_{\odot}$.  
Although part of the large dispersion in the GP values may be
due to the large uncertainties introduced by the assumptions
made on the gas mass fraction estimates, it seems clear that the
closed-box model do not reproduce the overall trend of the GPs. This is
especially remarkable in those GPs -- indicated by filled circles -- that 
show metallicities lower than the average given by the linear fit in  
Fig.~\ref{fig:2}. 
Fig.~\ref{fig:4} suggest that more sophisticated models, including outflows 
of enriched gas and inflows of pristine gas, should be needed to better address
the chemical evolution of the GPs.

\section{Discussion}
\label{sec:3}
  
Feedback processes associated with the vigorous star formation taking
place in the GPs can be invoked to explain the above results.  Large
amounts of supernovae (SNe) are expected to be produced by the
starbursts on very short time scales. The enriched material produced
by these SNe might escape from the galaxy, overcoming a relatively
weak potential, diluting metal abundances and explaining the offset in
the MZR.  Simulations (e.g., [16]) and observations at low (e.g., [8])
and high redshift (e.g., [9]), have invoked SNe feedback to explain
the shape of the MZR in low-mass star-forming galaxies.  In contrast,
analytical models by [17] show that any subsequent star formation
event will remove a MZR offset, unless galaxies have an inefficient
star formation.

A different view of the feedback processes comes from some analytical
models and simulations (e.g., [18], [19]). They show that under some
hydrodynamical conditions which are expected to be present in
extremely compact, high specific SFR objects such as the GPs, the
material ejected and accelerated by the SNe can cool down efficiently
and be reinserted into the star-forming regions. This way, the enriched
material never escapes from the galaxy and is reprocessed by successive
generations of stars until most of the gas is transformed into new stars. 
This is the so-called ``positive feedback'', a process that
would act in extremely short time scales and thus requires a
high star formation efficiency. Positive feedback would
explain the offset of the GPs in the MZR. However, some chemical evolution
models (e.g., [20]) have shown that higher star formation
efficiencies produce lower values in N/O at a given metallicity in the
range 12$+\log($O/H$) < 8.4$, which is the opposite to the observed
trend on the N/O -- O/H diagram (Fig.~\ref{fig:1}).

Given the similarities found between the GPs and starbursts at higher 
redshift, i.e., LBGs, an inflow scenario is especially interesting 
since it has recently been suggested that at redshift $z \sim 2-3$
(e.g., asterisks in Fig.~\ref{fig:2}) accretion of cold gas may be
the main driver of star formation and stellar mass growth 
(e.g., [21], [22], [23]). 
Inflow of metal-poor gas, either from the outskirts of the galaxy or
beyond, that can dilute metals in the galaxy centers [16], explaining 
the offset to lower abundances in the MZR and in the N/O -- O/H diagram. 
In the models by [16], the dilution
of metals due to an inflow can be restored by the effects of star
formation depending on the dilution-to-dynamical timescale ratio.
Since this ratio depends inversely with galaxy radius, galaxies with
smaller radius, such as the GPs, may be expected to take longer time
to enhance their oxygen abundances to the values expected from the
MZR (predicted by the closed-box model).  

Additional support to this scenario comes from the morphological 
analysis of few objects from spatially resolved HST images ([4,8]). 
These GPs show clumpy and disturbed morphologies probably associated with 
interactions/mergers.
Recent accretion of cold gas can be due to interactions with
small gas-rich galaxies or even large gas clouds. This eventually
increases the gas surface density and consequently enhances the star
formation. In models by [24] the rapid
decrease of the oxygen abundance during an episode of massive and
rapid accretion of metal-poor gas is followed by a slower evolution
which leads to the closed-box relation, thus forming a loop in the N/O
-- O/H diagram. Under this model, the GPs would be at the first stages
after infall triggers enhanced star formation, being very bright and
moving slowly from the left to the right in Fig.~\ref{fig:1}, until
the galaxies will finally arrive to the closed-box position. In order 
to reproduce the observed trends, further work might use these and 
other suitable models to quantify the relative importance of the 
accretion rates and/or the rate at which the galaxies expel enriched gas.

\section{Conclusions}
Green pea galaxies are a genuine population of metal-poor, luminous
and very compact starburst galaxies.  Results by [4], [5], and [6] 
permitted first clues to the evolutionary status of GPs. 
However, these galaxies and other similar samples in the local Universe 
should be analyzed in more detail to gain further insights into their nature,
buildup process and evolutionary pathways.  In particular, the direct
assessment of the gas fraction, the refinement of the stellar mass
estimations and their star formation history, as well as a proper 
comparison with detailed models of chemical evolution, are important 
issues to be addressed.

Recent deep and high signal-to-noise imaging and spectroscopic
observations with OSIRIS at the 10-m. Gran Telescopio Canarias (GTC)
(Amor\'in et al. 2011, in prep) will provide new insights on the
evolutionary state of the GPs.  In particular, we will be able to see
whether the GPs show an extended, old stellar population underlying the 
young burst, like those typically dominant in terms of stellar mass in 
most BCGs (e.g., [25], [26], [27]).  The age, metallicity and mass of
the old and young stellar populations will be analyzed in more detail 
by fitting population and evolutionary synthesis models to the observed
spectra. 

\begin{acknowledgement}
We are very greatful to Polychronis Papaderos, Gerhard Hensler, and 
Simone Recchi for valuable comments on this manuscript.
 This work has been funded by grants AYA2007-67965-C03-02, and
 CSD2006-00070: First Science with the GTC ({\url
   http://www.iac.es/consolider-ingenio-gtc/}) of the
 Consolider-Ingenio 2010 Program, by the Spanish MICINN.
\end{acknowledgement}

\end{document}